\documentstyle[psfig,spie]{article}

\begin{document}

\author{I.I. Mazin \\
Geophysical Laboratory and Center for High-Pressure Research\\
[12pt] Carnegie Institution of Washington, 5251 Broad Branch Rd. N.W.,
Washington,D.C. 20005-1305\\
[12pt] A. Golubov\\
[12pt] ISI, Research Centre J\"{u}lich (KFA), FRG\\
[12pt] and Institute for Solid State Physics, Chernogolovka, Russia}
\title{Impurity scattering in highly anisotropic superconductors and inerband sign
reversal of the order parameter.}
\date{January 1996}
\maketitle

\begin{abstract}
We discuss various mechanisms that can lead to interband sign reversal of
the order parameter in a multiband superconductor. In particular, we
generalize Abrikosov-Gor'kov solution of the problem of weakly coupled
superconductor with magnetic and nonmagnetic impurities on the case of
arbitary order parameter anisotropy, including extreme cases as $d-$pairing
or interband sign reversal of the order parameter, and show that interband
scattering by magnetic impurities can stabilize an interband sign-reversal
state. We discuss a possibility of such state in YBa$_2$Cu$_3$O$_7$ in the
context of various experiments: Josephson tunneling, neutron scattering,
isotope effect measurements. \ 
\end{abstract}

\date{}

1. In the last years new experiments with high-temperature superconductors
revived the interest to the strongly anisotropic superconductivity. Under
this name we mean superconductors where the order parameter $\Delta $ is
strongly different on different parts of the Fermi surface. It includes the
cases when strong anisotropy of $\Delta $ is implied by symmetry of the
superconducting state, like $d$-wave pairing, one-band superconductors with
a strong anisotropy, but the same symmetry as in the normal state (so-called
extended $s$ pairing), or multiband superconductors with substantially
different order parameters in different bands. It does not include
superconductors with anisotropic electronic structure
 which does not result in anisotropic order parameter.

In this paper we shall concentrate on the case of multiband superconductors,
more precisely on the case of extreme interband 
anisotropy when the sign of the order parameter is different in different
bands. We shall discuss various conditions which can lead to such a state.
First, we shall address the question whether a combination of phonon-induced
attraction and homogeneous Coulomb repulsion may lead to interband sign
reversal of the order parameter. Second, we shall derive a generalization of
the Abrikosov-Gor'kov formula for superconductivity in dirty superconductors%
\cite{AG} for arbitrary anisotropy of the order parameter, and will show
that impurity scattering does not necessarily reduce the anisotropy of the
superconducting state. Moreover, we shall point out cases when impurity
scattering can stabilize a solution with an interband sign reversal.
Finally, we will describe a model when pairing appears due to
electron-paramagnon scattering and the resulting state has an interband sign
reversal of the order parameter. Finally, we will discuss various
experiments in YBa$_2$Cu$_3$O$_7,$ and show how they can be explained with
the interband sign reversal concept.

2. The extension of the BCS theory for two or more superconducting bands was
first worked out by Suhl, Matthias, and Walker\cite{MSW} and independently
by Moskalenko \cite{Mosc}, and later elaborated on by many. It was realized%
\cite{df-met} that the fact that several bands cross the Fermi level is not
sufficient to have considerable many-band effects in superconductivity. Only
when the bands in question have a very different physical origin, can a
substantial effect appear. It has also been realized that the anisotropy, if
any, appears already in the weak coupling (BCS) solution, while the proper
strong coupling treatment adds
no qualitative features.

Many high-$T_c$ cuprates not only have multisheet Fermi surfaces, but the
actual bands
at the Fermi level are qualitatively different. In particular, YBa$_2$Cu$_3$O%
$_7$ is known to have four sheets of the Fermi surface, all four having
different physical origins\cite{Kanazawa}: One is formed by the chain $%
pd\sigma $ states 
(seen by positron annihilation), another is an apical oxygen band (seen in
de Haas-van Alphen experiments), and the last two are bonding and
antibonding combinations of the two $pd\sigma $ plane bands (seen by
angular-resolved photoemission). Basing on the richness of the band
structure of YBa$_2$Cu$_3$O$_7,$ several groups pointed out that at least
the two-band\cite{kresin}, or probably the whole four-band\cite{eilat,genzel}%
, picture should be used to describe superconductivity in this system.
Various experiments have been interpreted as indicating two or more
different superconducting gaps.

We shall now remind the basic equations of the multiband BCS theory\cite
{MSW,df-met}: The Hamiltonian has the following form: 
\[
H=\sum_{i,k\sigma \alpha }\epsilon _{i,k}c_{i,k\alpha }^{*}c_{i,k\alpha
}-\sum_{ij,kk^{\prime }\alpha \beta }\frac{g_{ij}}2c_{i,k\alpha
}^{*}c_{j,k^{\prime }\alpha }c_{i,-k\beta }^{*}c_{j,-k^{\prime }\beta } 
\]
where $\epsilon _{i,k}$ is the kinetic energy in the $i$-th band, $%
c_{i,k\sigma }^{*}$ and $c_{i,k\alpha }$ are corresponding creation and
annihilation operator, and $g_{ij}$ is the averaged pairing potential.

The order parameter $\Delta $ on the $i$-th sheet of the Fermi surface is
given by the equation 
\begin{equation}
\Delta _i=\sum_j\Lambda _{ij}\Delta _j\int_0^{\omega _D}dE\frac{\tanh (\sqrt{%
E^2+\Delta _j^2}/2k_BT)}{\sqrt{E^2+\Delta _j^2}},  \label{gaps}
\end{equation}
if the cut-off frequency $\omega _D$ is assumed to be the same for all
sheets. $T_c$ is defined in the usual way by the effective coupling
constant, $\log (2\gamma ^{*}\omega _D/\pi T_c)=1/\lambda _{eff},$ $\gamma
^{*}\simeq 1.78.$ The effective coupling constant $\lambda _{eff}$ in this
case is simply the maximal eigenvalue $\lambda _{\max }$ of the matrix $%
\Lambda _{ij}=g_{ij}N_j,$ where $N_j$ is the density of states at the Fermi
level (per spin) in the $j$-th band. $\Lambda _{ij}$ plays the role of the
coupling constant $\lambda $ in the one-band BCS theory. Note that
conventional (isotropic) $\lambda $ is also defined in terms of $\Lambda
_{ij}$: $\lambda =$ $\sum_{ij}\Lambda _{ij}N_i/N=\sum_i\lambda _iN_i/N$,
where the mass renormalization for the $i$-th band, as measured, for
instance, in de Haas-van Alphen experiments, is $\lambda _i=\sum_j\Lambda
_{ij}$, and $N=\sum_iN_i$. Obviously $\lambda _{eff}\geq $ $\lambda ,$ which
means that due to larger variational freedom $T_c$ in the multiband theory
is always larger than in the one-band theory. The two are equal in isotropic
case, i.e. when $g_{ij}$ does not depend on $i,j$. An instructive example of
the opposite case is the two-band model with $\Lambda _{11}=\Lambda
_{22}=\Lambda >0$, $\Lambda $$_{12}=\Lambda _{21}=-\Lambda .$ Then $\lambda
=0$, while $\lambda _{eff}=2\Lambda $. Note that the last value is the same
as when $\Lambda $$_{12}=\Lambda _{21}=\Lambda $. The physical reason is
that although there is no solution of Eq.\ref{gaps} with $\Delta _1>0$, $%
\Delta _2>0$, there is an obvious solution with $\Delta _1=-\Delta _2\neq 0.$
Near $T_c,$ the solution of Eq.\ref{gaps} is $\Delta _2/\Delta _1=(\lambda
_{eff}-\Lambda _{11})/\Lambda _{12},$ demonstrating directly that the sign
reversal of the order parameter, $\Delta _2/\Delta _1,$ takes place\ {\it %
when nondiagonal matrix elements\ }$\Lambda _{12}${\it \ and\ }$\Lambda $$%
_{21}${\it \ are negative}. One can easily check that Eq.\ref{gaps} may have
a superconducting solution even for all $g_{ij}<0$, {\it i.e., }when no
attractive interaction is present in the system. The condition for that is $%
|g_{12}|>(|g_1|N_1^2+|g_2|N_2^2)/2N_1N_2$. This is similar to the well-known
fact that in a system with repulsion the superconductivity with higher
angular momenta ($p,$ $d$) is possible, because of the sign reversal of the
order parameter. The main difference is that in the example above the
symmetry of the superconducting state is the same as of the normal state.
Below we shall demonstrate that even a fully attractive interaction $%
g_{ij}\geq 0$ can lead to the sign reversal if (a) interband pairing
interaction is weaker than Coulomb pseudopotential, or 
 (b) there is strong
interband scattering by magnetic impurities.

\begin{figure}[tbp]
\centerline{\psfig{file=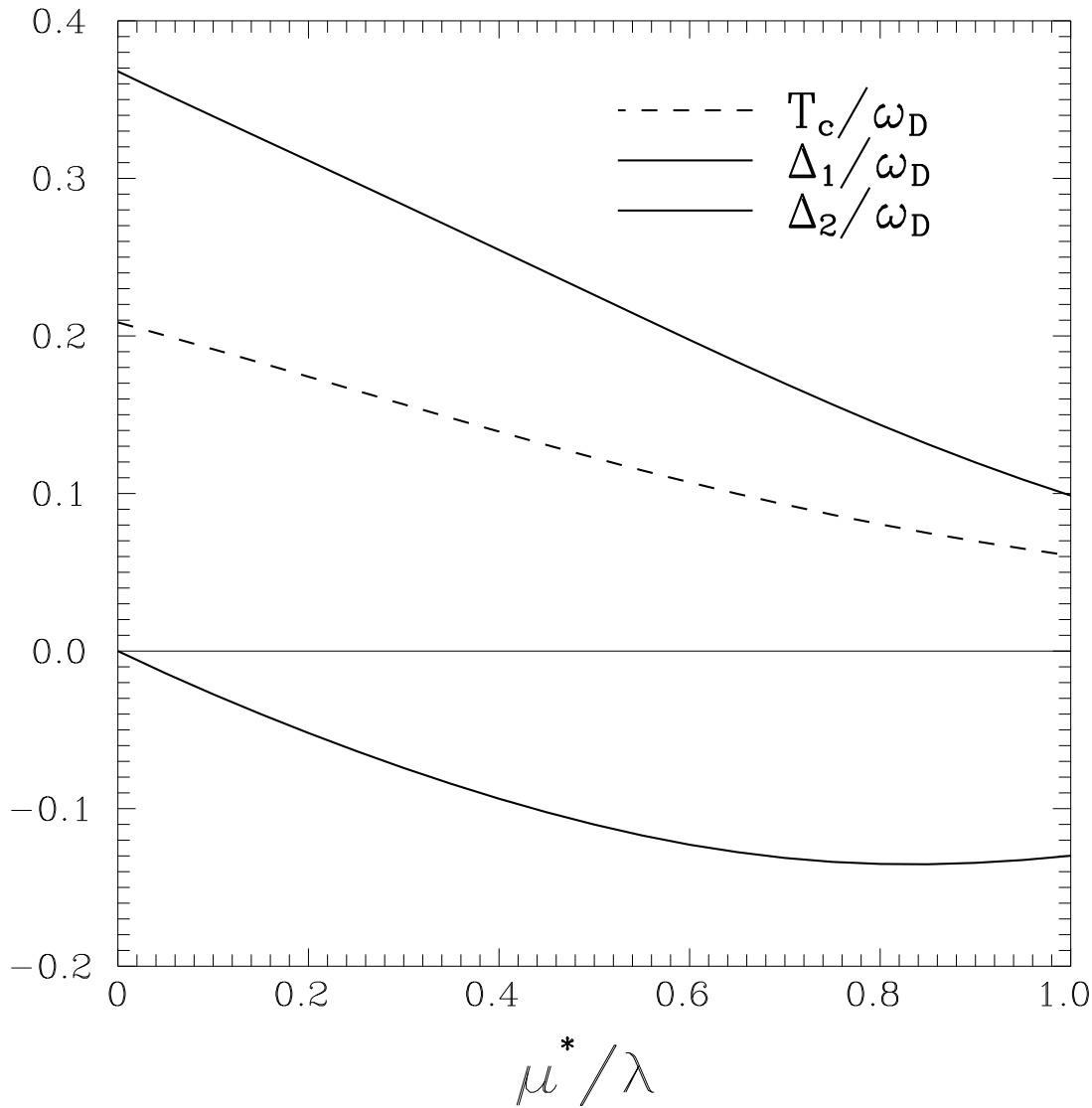,width=0.95\linewidth}}
\caption{Critical temperature and superconducting gaps in a system with
induced superconductivity in the second band, as a function of Coulmb
pseudopotential. }
\label{fig1}
\end{figure}

If $g$'s are electron-phonon pairing potentials, then Eq.\ref{gaps} should
be corrected for a Coulomb repulsion, which can be readily done\cite{df-met}
by substituting $g_{ij}\longrightarrow g_{ij}-U_{ij}^{*}\approx $ $%
g_{ij}-U^{*}$, where the effective Coulomb repulsion $U^{*}$ is
logarithmically renormalized in the same way as in one-band
superconductivity theory ($U^{*}$ is assumed to be independent on $i,j$). A
direct consequence of that is that if the interband electron-phonon coupling
is weak, the situation with a negative gap, $g_{ij}-U^{*}<0$, can easily be
realized because of the interband repulsion. We illustrate that by numerical
calculations presented in Fig.\ref{fig1}. In these 
 calculations the following
parameters had been used: $g_{12}=g_{22}=0$, $N_1=4N_2$, and $%
g_{11}=N_1^{-1} $ so that to have $\lambda =1$. 

Several facts draw attention: First, in this model $T_c$ decreases with the
increase of $\mu ^{*}=U^{*}N$ substantially slower than in a one-band case
when $T_c\rightarrow 0$ when $\mu ^{*}\rightarrow \lambda $. Second, the
order parameter induced in the second band (not superconducting by itself)
is always negative; Its absolute value reaches maximum when $|\Delta
_1|=|\Delta _2|,$ {\it i.e}., at $U^{*}=g_{pp}N_p/2(N_p-N_c)$. Further below
we shall discuss the relevance of this situation to YBa$_2$Cu$_3$O$_7.$

3. Now we shall analyze the influence of impurity scattering on the sign and
anisotropy of the order parameter. Following the standard way of including
the impurity scattering in the BCS theory\cite{AG}, one writes the equations
for the renormalized frequency $\tilde{\omega}_n$ and order parameter $%
\tilde{\Delta}_n$ ($n$ is the Matsubara index), which completely define the
superconductive properties of the system: 
\begin{eqnarray}
\hbar \tilde{\omega}_{Jn} &=&\hbar \omega _n+\sum_{J^{\prime }}\frac{\hbar ^2%
\tilde{\omega}_{J^{\prime }n}}{2Q_{J^{\prime }n}}(\gamma _{JJ^{\prime
}}+\gamma _{JJ^{\prime }}^s)  \nonumber \\
\tilde{\Delta}_{Jn} &=&\Delta _J+\sum_{J^{\prime }}\frac{\hbar ^2\tilde{%
\Delta}_{J^{\prime }n}}{2Q_{J^{\prime }n}}(\gamma _{JJ^{\prime }}-\gamma
_{JJ^{\prime }}^s)  \label{AG} \\
\Delta _J &=&\pi T\sum_{J^{\prime },n}^{\left| \omega _n\right| <\omega
_D}\Lambda _{JJ^{\prime }}\tilde{\Delta}_{J^{\prime }n}/{Q}_{J^{\prime }n}. 
\nonumber  
\end{eqnarray}
The only difference from a specific case of multiband superconductivity
(i.e., interband anisotropy), considered in a number of papers (see, e.g.
Ref. \cite{entel}) is that instead of the band indices in \ref{AG} appear
indices $J,J^{\prime }$, labeling Allen's Fermi surface harmonics\cite{allen}%
. Note that for a spherical (cylindrical) Fermi surface Allen's harmonics
reduce to spherical harmonics and $J$ to $\{L,m\}$ or just $m$. For the
interband anisotropy without intraband anisotropy $J$ can be chosen to be
the band index (so-called disjoint representation\cite{allen}). Other
notations in Eq.\ref{AG} have their usual meaning: $\omega _n=(2n+1)\pi T$, $%
Q_{Jn}=\sqrt{\tilde{\omega}_{Jn}^2+\tilde{\Delta}_{Jn}^2}$, $\gamma
_{JJ^{\prime }}$ is the scattering rate matrix by
nonmagnetic impurities, and $\gamma $$_{JJ^{\prime }}^s$ is the same for
magnetic impurities. Coupling matrix $\Lambda $ is defined in the same way
as Allen's
matrix $\lambda _{JJ^{\prime }}$ \cite{allen}, $\Lambda _{JJ^{\prime
}}=V_{JJ^{\prime }}^{{\rm pairing}}N_{JJ^{\prime }}(0)$.

Linearizing Eqs. \ref{AG} with respect to $\Delta $ (an approximation valid
near $T_c$), we get: 
\[
\Delta _{Jn}^{\prime }(1+G_J/2\omega _n)=\Delta _J+\sum_{J^{\prime }}\Delta
_{J^{\prime }n}^{\prime }\Gamma _{JJ^{\prime }}/2\omega _n\Rightarrow \Delta
_{Jn}^{\prime }=\sum_{J^{\prime }}\Delta _{J^{\prime }}(\delta _{JJ^{\prime
}}+g_{JJ^{\prime }}/2\omega _n)^{-1}, 
\]
where we have defined 
\begin{eqnarray*}
Z_{Jn} &=&1+G_J/2\omega _n;\,\,\,\,\tilde{\Delta}_{Jn}=\Delta
_J+\sum_{J^{\prime }}\frac{\tilde{\Delta}_{J^{\prime }n}}{2\omega
_nZ_{J^{\prime }n}}\Gamma _{JJ^{\prime }} \\
\Delta _{Jn}^{\prime } &=&\tilde{\Delta}_{Jn}/Z_{Jn};\,\,\,\,G_J=\sum_{J^{%
\prime }}\gamma _{JJ^{\prime }};\,\,\,\,Z_{Jn}=\tilde{\omega}_{Jn}/\omega _n
\\
\Gamma _{JJ^{\prime }} &=&\gamma _{JJ^{\prime }}-\gamma _{JJ^{\prime
}}^s;\,\,\,\,g_{JJ^{\prime }}=\delta _{JJ^{\prime }}G_J-\Gamma _{JJ^{\prime
}}.
\end{eqnarray*}

From this immediately follows that 
\[
\Delta _J=2\pi T\sum_{J^{\prime }n}^{\omega _n<\omega _D}\Lambda
_{JJ^{\prime }}\omega _n^{-1}\sum_k\Delta _k(\delta _{kJ^{\prime
}}+g_{kJ^{\prime }}/2\omega _n)^{-1},
\]
which after a usual trick with subtracting the clean limit, $g=0$, and
extending summation to infinity (a useful matrix formula is $(${\sf %
\^{I}+\^{A}}$)^{-1}=${\sf \^{I}-\^{A}(\^{I}+\^{A})}$^{-1}$), gives 
\begin{equation}
\Delta _J=\sum_{J^{\prime }k}\Lambda _{JJ^{\prime }}[L\delta _{J^{\prime
}k}-X_{J^{\prime }k}]\Delta _k;\,\,\,X_{JJ^{\prime }}=\sum_kg_{kJ^{\prime
}}/2\sum_n\omega _n^{-1}(\omega _n\delta _{Jk}+g_{Jk}/2)^{-1},  \label{X}
\end{equation}
where $L=\ln (2\gamma ^{*}\omega _D/\pi T_c).$ By introducing the
eigensystem of $g$, $g_{JJ^{\prime }}=\sum_kR_{Jk}^{-1}d_kR_{kJ^{\prime }},$
we can express $X$ in terms of the difference between the two incomplete
gamma-functions: 
\[
\,\,X_{JJ^{\prime }}=\sum_kR_{Jk}^{-1}\sum_n\omega _n^{-1}d_k(\omega
_n+d_k/2)^{-1}R_{kJ^{\prime }}=\sum_kR_{Jk}^{-1}\chi (d_k)R_{kJ^{\prime }},
\]
with $\chi (x)=\psi (1/2+x/2\pi )-\psi (1/2)$, which is the standard
definition of the matrix function ${\sf \hat{X}=}\chi ({\sf \hat{g})}.$

This result is analogous to the classical one of Abrikosov and Gor'kov\cite
{AG}, but includes arbitrary anisotropy. Now solving (\ref{X}) for $L$, we
find: 
\begin{equation}
\Delta _J=\sum_k(\Lambda _{Jk}^{-1}+X_{Jk})^{-1}L\Delta _k,  \label{last}
\end{equation}
which means that now $T_c$ is defined by the maximal eigenvalue of the {\it %
effective }matrix $\Lambda _{eff}=(\Lambda ^{-1}+X)^{-1}.$ As can be seen
immediately from Eqs.(\ref{X}--\ref{last}) and the definition of $%
g_{JJ^{\prime }}$, diagonal nonmagnetic scattering rates $\gamma _{ii}$ have
dropped out from Eq.(\ref{last}). This is the manifestation of Anderson
theorem for an anisotropic case: such scattering does not influence $T_c$
(in a considered Born limit), as in isotropic superconductors. As will be
discussed below, this argument works only for diagonal in $J,J^{\prime }$
non-magnetic scattering, while all other are, in principle, pair-breaking.

In the second order in $\Lambda $ (assuming that $\Lambda X$ is small), 
\begin{equation}
\Lambda _{eff}=\Lambda -\Lambda X\Lambda .
\end{equation}
If we recall that $\Delta $ forms the eigenvector of $\Lambda $ for the
maximal eigenvalue $\lambda _{eff}$, we can immediately write the
lowest-order correction to $\lambda _{eff}$:

\begin{equation}
\,\delta \lambda _{eff}=-\sum_{JJ^{\prime }}\Delta _JX_{JJ^{\prime }}\Delta
_{J^{\prime }}/L^2\sum_J\Delta _J^2
\end{equation}
$\,$

It is illustrative to consider explicitly a two-band case for weak
scattering ($\gamma _{ij},\gamma _{ij}^s\ll T_c)$ ; the effective matrix is
then given by:

\begin{equation}
\hat{\Lambda}_{eff}=\hat{\Lambda}-\frac \pi {8T_{c0}}\hat{\Lambda}\cdot
\left( 
\begin{array}{ll}
2\gamma _{11}^s+\gamma _{12}^s+\gamma _{12} & \gamma _{12}^s-\gamma _{12} \\ 
\gamma _{21}^s-\gamma _{21} & 2\gamma _{22}^s+\gamma _{21}^s+\gamma _{21}
\end{array}
\right) \cdot \hat{\Lambda}.
\end{equation}

When all $\Lambda $'s are equal (isotropic case), the standard
Abrikosov-Gor'kov result is recovered: $\delta \lambda \approx -\pi \lambda
^2(\gamma _{11}^s+\gamma _{12}^s+\gamma _{21}^s+\gamma _{22}^s)/8T_{c0}.$
The main point of the AG theory\cite{AG} is that $\gamma $$^s$ enters
equations for $\omega $ and $\Delta $ with opposite signs. That is why the
magnetic impurities appear to be pair-breakers, and the non-magnetic ones
not. The above solution shows explicitly that in anisotropic case of Eqs.\ref
{AG} only intraband non-magnetic scattering does not influence $T_c$ ($%
\gamma _{ii}$ drop out).

An interesting special case is $\Lambda _{12},\Lambda _{21}\ll \Lambda
_{11},\Lambda _{22}.$ Then in the effective $\Lambda $ matrix the
nondiagonal element $\Lambda _{12}^{eff}=\Lambda _{12}+\pi \Lambda
_{11}\Lambda _{22}(\gamma _{12}-\gamma _{12}^s)/8T_{c0}$ becomes negative,
if $\gamma _{12}^s-\gamma _{12}>8T_{c0}\Lambda _{12}/\Lambda _{11}\Lambda
_{22}$. As discussed above, this situation will lead to sign reversal of the
order parameter. One can say that, when attractive interband coupling is
relatively weak and the magnetic interband scattering is strong, the system
chooses to have two gaps of the opposite signs, losing in pairing energy,
but avoiding the pair-breaking due to interband scattering.

4. Maybe the most interesting example of an interband sign reversal of the
order parameter is superconductivity due to electron-paramagnon interaction,
in other words, superconductivity due to dynamic exchange of spin
fluctuations. For simplicity, we consider the case of singlet pairing. In
this case the pairing potential is $Vij,kl=\int {d{\bf R}d{\bf R}^{\prime }}%
\sum_{\alpha \beta \gamma \delta }\langle i\alpha |J({\bf r-R})\sigma
_{\alpha \beta }|j\beta \rangle \chi ({\bf R-R}^{\prime })\langle k\gamma |J(%
{\bf r-R})\sigma _{\gamma \delta }|l\delta \rangle $ where $J$ is exchange
interaction and $\chi =\langle {\bf S(R)S(R}^{\prime }{\bf )}\rangle $ is
spin-spin correlation function. Contrary to the electron-phonon interaction,
this interaction is repulsive (for triplet pairing, it would be attractive).
Thus it cannot lead to a superconducting solution where the order parameter
has always the same sign. A possible solution 
is the well-known $d$-wave state, where the order parameter changes sign
upon rotation by $\pi /2$ in the momentum
space. Another possibility is to have order parameter of different sign in
different bands. In this case the interaction must be small in the intraband
channels and strong in the interband channel. It looks on the first glance
to be an artificial, unnatural condition. However, in some cases it emerges
quite naturally. A good example is a bilayer with spin fluctuations
antiferromagnetically correlated between the layers. This system models some
superconducting cuprates, like YBa$_2$Cu$_3$O$_{7-\delta }$. If the
one-electron tunneling between the layers is larger than superconducting
gap, two bands are formed and well defined: bonding (symmetric) and
antibonding (antisymmetric). By symmetry, only electron states of different
parity can interact via exchange of antiferromagnetic (thus antisymmetric)
spin fluctuations. Spin-fluctuation induces 
superconductivity in YBa$_2$Cu$_3
$O$_{7-\delta }$ with sign reversal of the order parameter was
quantitatively investigated in Ref.\cite{licht}.

5. Now we shall briefly discuss some experimental implications of the
interband sign reversal of the order parameter, if it occurs in YBa$_2$Cu$_3$%
O$_7.$ It is believed that four different bands cross the Fermi level in YBa$%
_2$Cu$_3$O$_7$: two plane bands, which are bonding ($B$) and antibonding ($A 
$) combinations of the individual planes' states, the chain ($C)$ band, and
a small pocket formed mainly by apical oxygen states (which is not discussed
here). We adopt here the point of view that the superconductivity originates
in the plane bands, and is induced in the chain band by interband proximity
effect. Thus we shall consider a superconducting state which is
characterized by the order parameters of the opposite signs in the bonding
and antibonding bands. For the following discussion we shall also mention
that, according to calculations\cite{eilat}, the chain band is very light,
so that its contribution to the total density of states is small ($\sim 15$%
\%), while its contribution in the plasma frequency $\omega _{py}^2\propto
N(0)v_{Fy}^2$ is considerable ($\sim 50$\%). These finding are confirmed by
the experiment: Maximal Fermi velocity was calculated\cite{add} to be $\sim
6\times 10^7$ cm/s and corresponds to the point where the chain Fermi
surface crosses the $\Gamma -$Y line. This value agrees well with the Raman
experiments \cite{friedl}. Calculated plasma frequency anisotropy $\omega
_{py}^2/\omega _{px}^2\simeq 1.75$, as discussed in Ref.\cite{MD}, is in
agreement with the optical and transport measurements. Band $A$ is,
according to calculations, rather heavy, which also agrees with the
experiment \cite{shen,Gofron}. Band $B$ is light again. Both $A$ and $B$
bands are nearly tetragonal. Their relative contribution to the normal-state
transport is defined by the partial plasma frequencies (Table I).
Importantly, bands $A$ and $C$ at $q_z=0$ can cross by symmetry, for
instance they are degenerate with $\epsilon =E_F$ at {\bf q}=($\approx
0.8\pi /a,\approx 0.2\pi /B,0)$. For all $q_z\neq 0$ these bands hybridize.
This is the reason for YBCO being the most three-dimensional of all high-$%
T_c $ cuprates. An extremal orbit in $q_z=0$ plane, which appears because of
the $A-C$ hybridization, has been seen in de Haas-van Alphen experiments\cite
{dhva}.

Which of the three possible mechanisms (if any) can be operative in YBa$_2$Cu%
$_3$O$_7$ and produce interband sign reversal? It appears that any of the
three can. First, it is possible that the phonon-induced interaction is
stronger in the intraband than in the interband channel. For instance, one
of the key feature of the electronic structure of YBa$_2$Cu$_3$O$_7$ are
extended saddle points at van Hove singularities. It was shown that
extending of the saddle points is directly related to the geometry (warping)
of the CuO$_2$ planes\cite{TB8}. Saddle points appear only in the
antibonding band. Thus, one may expect the saddle points regions to interact
particularly strongly with the phonons of the even parity ({\it gerade}),
with correspondingly weaker intraband interaction. Second, the main type of
magnetic defects in YBa$_2$Cu$_3$O$_7$ are localized magnetic moments on Cu,
which appear near O vacancies and other imperfections. In view of very
strong interplane antiferromagnetic correlations of the Cu spins in
underdoped compounds, it seem plausible that each defect-related Cu moment
in the fully doped system induced magnetic moment at the nearest Cu site in
the next plane. Such pair of spins will have exactly that symmetry, that is
needed to stabilize the superconductivity with the opposite signs 
of the order parameter. Third, the popular model of the spin-fluctuation
superconductivity, which is usually associated with the $d$-pairing, in fact
always has another solution, namely that with the angular $s$-symmetry and
interband sign reversal\cite{licht}. Which of the two solutions is more
stable depends on whether or not the spin fluctuations are
antiferromagneticaly correlated between the planes. If they are, the second
solution is always more stable. Correspondingly, if one accepts the results
of the numerous neutron scattering experiments, which do show the interplane
correlation, and assumes the spin-fluctuation induced superconductivity, the
resulting state is one with the interband sign reversal. Note that this is
true both for the overdamped magnetic excitation deduced by Pines and
co-workers from the NMR data, and for the finite-energy paramagnon as
deduced from the neutron scattering.

Now we shall briefly discuss some experiments, probing the symmetry and/or
anisotropy of the superconducting state. It is instructive to single out
those few experiments that deal with the symmetry of the order parameter.
Those are Josephson tunneling experiments and neutron scattering
experiments. Both were discussed in detail, in the context of the interband
sign reversal, in Refs.\cite{jos} and \cite{yak}, respectively. Numerous
Josephson tunneling experiments, performed in the last two years, show that
the phases of the tunneling current along crystallographic $a$ and $b$
directions are shifted by $\pi .$ A natural interpretation is in the form of
the one-band $d_{x^2-y^2}$ state. As it was pointed out in Ref. \cite{jos},
because of the strong difference in effective masses, the chain band in YBa$%
_2$Cu$_3$O$_7$ contribute substantially into the transport in $b$ direction.
Thus, the relative phase of the tunneling currents in the interband sign
reversal model depends on whether the sign of the order parameter in the
chain band is the same as in the bonding band (which dominates in-plane
transport) or as in the antibonding band. In Ref. \cite{jos} it was argued
that the latter is true, since the chain and the antibonding band anticross
at the Fermi level. In this case reasonable numerical estimates lead to the
conclusion that the total tunneling current will have the desired phase
shift, unless the chain band gap is too small. The authors of Ref. \cite
{comb} made an opposite assumption, namely that the signs of the order
parameter in the chain and the antibonding band are different. This makes
the gap disappear near the band crossing and thus fits with various
experiments indicating vanishing minimal gap. However, it is more difficult
to explain the Josephson experiments in this model (one would have to assume
that both in the chain and in the bonding bands gaps are much smaller than
in the antibonding band).

While the interband sign reversal model 
offers an alternative explanation of the Josephson experiments, it appears
to be the only one able to explain the neutron scattering result in a
plausible manner. Currently there is a consensus in the neutron scattering
community that in fully doped YBa$_2$Cu$_3$O$_7$ a sharp peak in the
imaginary part of magnetic susceptibility $\chi ^{\prime \prime }$ appears
below $T_c$ at the plane wave vector {\bf q=(}$\frac \pi a,\frac \pi b)$,
which shows perfect antiferromagnetic correlations between the planes. The
most obvious effect of the onset of superconductivity on magnetic
susceptibility is the appearance of a peak in $\chi ^{\prime }$ at $\hbar
\omega =2\Delta $, which however is allowed only if the correspondent wave
vectors connects the parts of the Fermi surface with the opposite signs of
the order parameter. Another effect is the appearance of a peak in $\chi
^{\prime \prime }$ at $\hbar \omega =\Delta +\xi _{vH}$, where $\xi _{vH}$
is the position of the van Hove singularity. In the RPA approximation the
two peaks enhance each other, thus leading to a strong effect. However, the
only way to ensure that the peak has no ferromagnetic (between the planes)
component is to assume that there is no intraband sign reversal at {\bf q=(}$%
\frac \pi a,\frac \pi b)$, but only interband sign reversal.\cite{yak}

It is worth noting that intraband sign reversal does not exclude any angular
anisotropy of the order parameter. Moreover, the specific model considered
in Ref. \cite{licht} appeared to have rather strong angular variation of the
gap value, with the symmetry close to ($x^2-y^2)^2$. There are numerous
experiments pointing to large variation of the absolute value of the gap in
YBa$_2$Cu$_3$O$_7.$ As long as these experiments do not distinguish between $%
d-$ and 
strongly anisotropic $s-$wave states we do not discuss them here. A final
note concerns isotope effect, which is known to be small at the optimal
doping (the highest $T_c$), but increases rapidly when superconductivity is
suppressed by oxygen reduction, Pr doping, or other means. Such a behavior
is very typical for superconductivity of mixed phonon and non-phonon origin.
However, both components should work constructively, and not destructively.
For instance, in a regular electron-phonon isotropic superconductor spin
fluctuations, if any, are pair-breaking, and not pairing. One possibility to
rationalize the isotope affect is to say that the electronic component is
non-magnetic (like acoustic plasmons), or that the major process is
two-magnons exchange. Another possibility is to say that the phonon- and
spin fluctuation-induced interaction are separated in the phase space. For
instance, if (as it is usually assumed) spin fluctuations are operative near 
{\bf q=(}$\frac \pi a,\frac \pi b)$, but phonons only at small {\bf q}'s,
both can be pairing in the contest of the $d_{x^2-y^2}$ superconductivity.
Alternatively, if spin fluctuation are operative only in the interband
channel, as it follows from the neutron scattering experiments, and phonons
only in the intraband channel (in other words, only even [{\it gerade}]
phonons interact considerably with the electrons), the isotope effect trends
follow naturally.

\end{document}